\documentstyle[aps, prd, preprint, epsfig]{revtex} 
\tightenlines 
 
\def\stacksymbols #1#2#3#4{\def\theguybelow{#2} 
    \def\verticalposition{\lower#3pt} 
    \def\spacingwithinsymbol{\baselineskip0pt\lineskip#4pt} 
    \mathrel{\mathpalette\intermediary#1}} 
    \def\intermediary#1#2{\verticalposition\vbox 
{\spacingwithinsymbol 
    \everycr={}\tabskip0pt 
    \halign{$\mathsurround0pt#1\hfil##\hfil$\crcr#2\crcr 
    \theguybelow\crcr}}}

\def\eq{\begin{equation}} 
\def\en{\end{equation}} 
\newcommand{\beqn}{\begin{eqnarray}} 
\newcommand{\eeqn}{\end{eqnarray}} 
\def\part{\partial} 
 
\def\part{\partial}

\title{Renormalization and Essential Singularity} 
\author{Miyuki Nishikawa{\footnote{Electronic  
address:nisikawa@hep-th.phys.s.u-tokyo.ac.jp}}} 
\address{%
  Department of Physics, University of Tokyo, \\
  Bunkyo-ku, Tokyo, 113--0033, Japan
}
\makeindex 
  
\begin{document} 
\bibliographystyle{/users/lpthe/viallet/formats/perso} 
  
\begin{titlepage} 
\renewcommand{\thepage}{} 
\maketitle 
\hoffset 5truemm 
\begin{center} 
Physics Department, University of Tokyo\\
(submitted to PROGRESS OF THEORETICAL PHYSICS) 
\end{center} 
\hoffset 0truemm 
\vskip 2truecm 
\begin{abstract} 
In usual dimensional counting, momentum has dimension one. But 
a function $f(x)$, when differentiated $n$ times,  
does not always behave like one with its power smaller by $n$. 
This inevitable uncertainty may be essential in general theory of  
renormalization, including quantum gravity.   
As an example, we classify possible singularities of a potential for the  
Schr\"{o}dinger equation, assuming that the potential $V$  
has at least one $C^2$ class eigen function. The result  
crucially depends on the analytic property of the eigen function near its  
0 point. 
\end{abstract} 
\vfill 
\end{titlepage} 
\section{Introduction } 
 In usual dimensional counting, momentum has dimension one. But  
as a differential operator, momentum has no well-defined  
dimension unless the object it operates on is determined  
explicitly. For example, a function $f(x)$ with an essential  
singularity at $x=0, (f(0)=0)$, when differentiated $n$ times,  
does not always behave like one with its power smaller by $n$ in the  
neighborhood of $x=0$. This fact is essential throughout this  
paper, and so, though I will investigate on the general behavior  
of possible singularities of a potential by a simple model  
below, the discussion can be applied to a wide class of  
differential equations of physical importance (See section 5.) 
 
Quantum theory is accompanied with inevitable ambiguity beyond 
measurements. This work is motivated by an attempt to shed light on such  
ambiguity. For example, the analyticity of a wave function can not be  
determined by finite times of measurements. But if the wave function 
becomes very sharp at a point, momentum is dominated by the point. So,  
a theory should not be sensitive to such analyticity. Contrastingly,  
if a continuous wave function takes both plus and minus values, it  
must also takes 0, which is the qualitative fact not sensitive to the  
exact shape of a wave function. So in this paper I first try to distinguish  
these two kinds of singularities in section 2. Then I proceed 
to classify the most general types of possible singularities in section 
3. The mathematical results derived there are physically explained in 
section 4. Applications to realistic cases are discussed in section 5  
and 6.This paper has something to do with the theory of computability  
\cite{Comp ,Pen }.  
\section{Possibility of singularity and Domain of Definition } 
 For simplicity, let us consider a one-dimensional  
Schr\"{o}dinger equation,  
\beqn y''=Vy-Ey. \label{ko1 } \eeqn 
We will consider from now on if there exists a  
potential $V(x)$ for an arbitrary function $y(x)$ given as an eigen  
function.  
 
For this purpose, it is necessary to define a `potential'.  
The potential between two particles with a large  
enough distance $r$ from each other is  
\beqn \left \{ \begin{array}{l}\to  1/r \;\; 
\mbox{(for electroweak force and gravity)} \nonumber \\ 
 \to \mbox{not exactly known but gets large (proportional to $r$ in a  
nuclei) (for strong force)} \end{array} \right. \eeqn 
More precisely, the coupling constants depend on the energy of  
the system, or the power of $r$ changes. For very short  
distance, the potential is completely unknown because current  
theory does not work beyond the Planck scale and it is also  
impossible to measure the size of an electron or a quark without  
any experimental error. So, in this paper I will say a function  
$y(x)$ has a potential $V(x)$ iff $y$ is a $C^2$ class  
function defined in $(0, a)$ and if there exists a continuous  
function $V(x)$ defined in $(0, a)$ and a constant $E$ which  
satisfies (\ref{ko1 }). 
 
In fact, any $C^2$ class function $y$ satisfies (\ref{ko1 }) if  
we take 
\beqn V=y''/y, \;\; E=0. \label{ko2 } \eeqn 
Here the replacement of the constant $E\to E'$ is equivalent to  
$V\to V -E'+E$, so from now on we take $E=0$. There are 2 possible  
cases for (\ref{ko2 }) to have a singularity: 
 
(I) there exists a $x$ such that $y(x)=0$, $0<x<a$,  
 
(II) $y''$ does not converge (for $x\to +0$ or $x\to a-0$). 
 
\section{Classification of the Power of Possible Singularities } 
 Now let us move the possible singularity to $x=0$ by the  
redefinition of coordinates and consider the behavior of $V$ as  
$x\to +0$. Let $y(z)$ be an analytic continuation of $y(x)$ to the  
complex plane. 
 
\vspace{.2 in} 
\noindent (CASE 1) $y(z)$ has no essential singularity at $z=0$. 
 
\vspace{.1 in} 
\noindent (a) If $y(z)$ can be Laurent expanded around $z=0$ as  
\beqn y=\sum_{n=k }^\infty a_nz^n, \;\; a_k\neq 0, \label{ko4 }  
\eeqn 
then  
\beqn \frac{y'' }{y }=\frac{\sum_{n=k }^\infty  
a_nn(n-1)z^{n-2 }}{\sum_{n=k }^\infty a_nz^n} \to\left  
\{ \begin{array}{l}\frac{a_d }{a_k }d(d-1)z^{d-2-k }\;\; (0\leq  
k)  \\ 
k(k-1)z^{-2 }\;\; (k<0) \end{array} \right. , \label{ko5 } \eeqn 
with $d$ the lowest power such that $a_d\neq 0$ and $1<d$ (if  
there is no such $d$, $a_d=0$). 
 
\vspace{.1 in} 
\noindent (b) When we change the power of the finite number of  
terms in the type (a) expansion into an arbitrary real number,  
{\footnote{From now on, the expansion coefficients are all real  
except if mentioned, and the branch is chosen so that the 
function takes unique real value at $z\to +0$. More precisely, a  
branching point with the power of an irrational number is an  
essential singularity, but the difference is not important  
here\cite{uniq}.}}  
\beqn \frac{y'' }{y }\to\left\{ \begin{array}{l}\frac{a_d }{a_k } 
d(d-1)z^{d-2-k 
}\;\;\left (\begin{array}{l}\mbox{if}\;  y = a_0 +a_1z +a_dz^d\cdots  
\;\; \mbox{or} \\ 
y = a_1z +z_dz^d\cdots \end{array} \right )\\ 
k(k-1)z^{-2 }\;\; (k<0)\;\; \mbox{(otherwise)} \end{array} \right. , 
\label{ko6 } 
\eeqn 
where $a_d$ is the coefficient of the lowest power except for  
$0, 1$. 
Summarizing above, the powers $\nu$ where the potential can  
behave like  
$V\to x^\nu$ as $x\to +0$ are 
 
 for (I), $\nu =-2 \; ;\; -1\leq \nu$,  
 
 for (II), $-2\leq \nu <0$.

\vspace{.2 in} 
\noindent (CASE 2) $y(z)$ has an isolated essential  
singularity at $z=0$. 
In complex analysis, a sequence of points can converge to any  
value 
depending on its approach to an essential singularity (with  
infinite order) \cite{Ahlf }. But now that we deal with only the  
case along the real axis $z\to +0$, the limit is sometimes well  
defined. Let's study the following cases.  
 
\noindent (c)  
\beqn y=\sum_{n=l }^ka_n(\log z)^n, \;\; a_l\neq 0, \label{ko7 }  
\eeqn 
if the above expansion is possible, then 
\beqn \frac{y'' }{y }=\frac{\sum_{n=l }^kna_n\{ (n-1)(\log z)^{n- 
2 }-(\log  
z)^{n-1 }\} }{z^2\sum_{n=l }^ka_n(\log z)^n }\to\frac{-k }{z^2 
\log z } \label{ko8 } \;\; ,  
\eeqn 
where $\log z$ diverges as $z\to 0$, but for an arbitrary integer  
$n$, $z(\log z)^n$ tends to $0$. So we can regard $\log z$ as `an infinitely  
small negative power' $z^{-\epsilon }\;(\epsilon >0)$. Then we  
can generalize type (b) expansion by the replacement of the  
finite number of terms 
\beqn a_nz^n\to z^n\sum_{m=l_n }^{k_n }a_{mn }(\log z)^m\;\;  
(m\in R). \label{ko9 }  
\eeqn 
This has the effect of 
\beqn \left \{ \begin{array}{l}z^{d-2-k }\to z^{d-2-k }(\log z) 
^m\;\; 
       (m\in R)\\ z^{-2 }\to z^{-2 }/\log z \end{array} 
\right. \label{ko10 } \eeqn 
in (\ref{ko6 }), i.e.,  
\beqn \mbox{for (I), } \nu =-2 (+\epsilon )\; ;\;\; -1\leq \nu . 
\label{ko11 }  
\eeqn 
{\footnote{For a $C^2$ class function $y$, $-1-\epsilon$ is  
impossible. And for (II), the region of $\nu$ is invariant.}}  
Let's call this type of expansion type (c). We can  
define the index of power $k_y, \mu_y, \nu_y$ as $z\to 0$ for  
type (c)  expansions as follows: 
\beqn y\to z^{k_y }, \;\;\frac{y' }{y }\to z^{\mu_y }, \;\;\frac 
{y'' }{y }\to  
z^{\nu_y }.  \label{ko12 } \eeqn 
Type (c) property is invariant under finite times of  
summations, subtractions, and differentiations. 
  
\vspace{.1 in} 
\noindent (d) When we apply finite times of summations,  
subtractions, multiplications, divisions (by $\neq 0$),  
differentiations, and compositions (with the shape of $f(g(z)), \; 0 
\leq k_g, \; g( +0)= +0$ where $f, g$ are type (c) expansions),  
$k_y, \mu_y, \nu_y$ can also be defined. As an arbitrary type (d)  
expansion $f(z)$ has a countable number of terms and a  
nonzero `radius of convergence 
{\footnote{The meaning of this term is different from the usual one 
because $z=0$ can be a singularity point.}}  
' $r$ where the expansion  
converges for $0<|z|<r$, it can be written as  
\beqn f(z)=\sum_{n=0 }^\infty f_n \;\; . \label{ko13 } \eeqn 
As the `principal part' which satisfies $k_{f_n}<0$ consists  
of finite number of terms, a type (d) expansion diverges or  
converges monotonically as $z\to +0$, so enables  the expansion of  
(\ref{ko13 }) in the order of ascending powers. As the expansion  
is almost the same as that of type (c) (the only differences are the  
multiplications by $(\log z)^n$ for an infinite number of terms  
and the appearance of the terms like $\log (z\log z)$), the  
region of $\nu_y$ is invariant.  
 
\vspace{.1 in} 
\noindent (e) When the following expansion is possible (type (e)): 
$y=\pm e^{f(z) }$, where $f$ is a type (d) expansion. We can  
define the finite values $\mu_y, \nu_y$ by  
\beqn \left \{ \begin{array}{l}\frac{y' }{y }=f'\to z^{\mu_y },  
\;\; \mu_y=k_f +\mu_f , 
\nonumber \\ 
\frac{y'' }{y }=f'^2 +f''\to z^{\nu_y }, \;\;\nu_y\geq {\mbox{min}}(2k_f  
+2\mu_f, \;  
k_f +\nu_f) .\end{array} \right. \label{ko14 } \eeqn 
Let us consider the region of $\nu_y$. For $k_f\geq  
0$ it is the same as for the type (d). For  
\beqn y=e^{az^k }, \;\; a, \; k\in R, \;\; k\leq 0 \label{ko15 }  
\eeqn 
satisfies 
\beqn  
\frac{y'' }{y }=a^2k^2z^{2k-2 } +ak(k-1)z^{k-2 }\to\left\{  
\begin{array}{l} 
z^{-2\pm\epsilon }\;\;(k=-\epsilon )\nonumber \\ 
a^2k^2z^{2k-2 }\;\; (k<0) \end{array}, \right. \label{ko16 }  
\eeqn 
combination with type (c) case leads to the region  
of $\nu_y$ being:  
 
for (I), $\nu_y\leq -2 +\epsilon \; ; \;\; -1\leq \nu_y$, 
 
for (II), an arbitrary negative number.  
 
\noindent Let us then consider if we can fill the remaining `window'  
of the region of $\nu_y$ for (I),  
 
$-2 +\epsilon <\nu_y<-1$. 
 
\vspace{.1 in} 
\noindent (f) When we can write $y=f_0 +\sum_{n=1 }^{m }(\pm )e^ 
{f_n }$, where $f_n$ is of type (c), $k_{f_n }<0$, and ($\pm$)  
takes each of the signatures $ +-$. 
We can assume that each terms in $\sum$ are ordered in the  
increasing absolute values for $z\to +0$. Then  
\beqn e^{az^k }\to\left \{\begin{array}{l}z^0\;\;(k\geq 0, \;  
a=0)\nonumber \\ 
0\;\;\; (k<0, \; a<0)\nonumber \\ 
\infty\;\; (k<0, \; a>0) \end{array}\right. \label{ko18 } \eeqn  
and as $y\to 0$ for (I),  
 
\beqn y=\left (\sum_{n=0 }^\infty\sum_{m=l_n }^{m_n }a_{nm }z^n 
(\log  
z)^m\right ) +\sum_{n=1 }^l(\pm ) 
e^{\sum_{i=k_n }^\infty\sum_{j=l_{ni } }^{k_{ni } }a_{nij }z^i(\log z)^j 
} 
. \label 
{ko19 } \eeqn  
If the second term sum at the R.H.S. is not 0, we can write 
 
\beqn k_1<\cdots <k_l<0, \;\; a_{nk_nk_{ni } }<0. \label{ko20 }  
\eeqn  
As $y$ is of $C^2$ class, the first term can be written  
as  
\beqn (\;\; )=a_{10 }z +\sum_{n=2 }^\infty\sum_{m=l_n }^{m_n } 
\cdots , \;\; m_2=0.   
\label{ko21 } \eeqn  
As 
\beqn y''\to \left \{  
{\begin{array}{l}(z^n(\log z)^m)'\to z^{n-m\epsilon -2 }  
\;\;\left (\begin{array}{l}  
\mbox{The term such that } \\  
\mbox{$n-m\epsilon$ is the smallest} \end{array}\right )  
\;\; (^\exists a_{nm} \neq 0) \\ 
\left [\{ a_{nk_nk_{ni } }z^{k_n }(\log z)^{k_{ni} }\}  
'^2+\{ a_{nk_nk_{ni } }z^{k_n }(\log z)^{k_{ni} }\}  
'' \right ] \\ 
\hspace{49mm} 
\times e^{\sum_{i=k_n }^\infty\sum_{j=l_{ni }}^{k_{ni } } 
a_{nij }z^i(\log z)^j}\;\;\; ( ^\forall a_{nm }=0) \end{array}} 
\right.\label{ko22 } \eeqn  
for  $z\to +0$,  
\beqn \frac{y'' }{y }\to \left \{ \begin{array}{l} 
z^{n-m\epsilon -3 }\;\;  
(a_{10 }\neq 0\; \mbox{and}\;  ^\exists a_{nm}\neq 0)\\ 
z^{2k_n-2k_{ni }\epsilon -2 }e^{a_{nk_nk_{ni } }z^{k_n } 
(\log z)^{k_{ni } }}\to 0\;\;  
(a_{10 }\neq 0\; \mbox{and}\; ^\forall a_{nm }=0) \\ 
z^{-2 }\;\; (a_{10 }= 0\; \mbox{and}\;  ^\exists a_{nm }\neq 0) \\ 
z^{2k_n-2k_{ni }\epsilon-2 }\;\;(a_{10 }= 0\; \mbox{and}\; ^\forall  
a_{nm }=0)  
\end{array} \right. . \label{ko23 } \eeqn  
The possible values of $\nu_y$ for (I) remain the same: 
$\nu_y\leq -2 +\epsilon \; ; \;\; -1\leq \nu_y$. 
 
\vspace{.1 in} 
\noindent (g) Whole of the expansions obtained from type (f)  
expansions by finite times of summations, subtractions,  
multiplications, divisions (by $\neq 0$), differentiations, and  
compositions (with the shape of $f(g(z)), \; 0\leq k_g, \; g( +0)=  
+0$ where $f, g$ are type (f) expansions). 
 
This type of expansion is very complicated compared to  
an ordinary Laurent expansion, but in any case has a countable  
number of terms and a nonzero `radius of convergence' $r$  
{\footnote{Of course, the meaning is different from the usual  
one. }} 
where $y$ is analytic for $0<|z|<r$. This can also be ordered  
partially in the ascending powers and we can write the first  
term explicitly, and so monotonically diverges or converges but  
never oscillates as  $z\to +0$. Its general shape is the whole  
sum  
\beqn (1)_i +(2)_j +\cdots +(m)_k \; , \label{ko25 } \eeqn  
where 
\beqn (1)_i \;\; & := & \mbox{\huge (}\sum_{n\in \{ n\}_i 
}^\infty\sum_{m_1,\cdots , m_{d_i }=-\infty }^{m_{i1 }, \cdots , m_{id_i 
} }a_{inm_1\cdots m_{d_i } }z^n(-\log z)^{m_1 }(-\log (-z/\log 
z))^{m_2}\nonumber \\ &    &  \hspace{4.2 cm}\cdots (-\log (-z/(-\log  
(-z/\log\cdots z))))^{m_{d_i } } \mbox{\huge )}_i, \nonumber \\ 
(2)_{\pm j } & := & \sum_{i\in \{ i\}_j }(\pm )e^{\pm (1)_i },  
\nonumber \\ 
(3)_{\pm k } & := & \sum_{j\in \{ j\}_k }(\pm )e^{\pm (2)_j },  
\nonumber \\ 
\vdots \label{ko24 } \eeqn 
Here the ($\pm$) in front of $e$ takes each of the  
signatures depending on each $i$ (or $j, k, \cdots $), while the  
$\pm$ on the shoulder of $e$ and in front of $j, k, \cdots $  
takes the signature such that the coefficient of the first term  
in $\sum$ is of the same signature as $j$ after choosing the  
signatures. Each term is ordered in partially ascending powers 
with regards for any sums. The sum with index  $n$ is performed  
according to the monotonically non-decreasing sequence of real  
numbers $\{ n_i\}\; (-\infty <n_i)$ depending on $i$. In the  
same manner, the sum with index  $i, j, \cdots $ is performed  
according to the finite, monotonically non-decreasing sequence $\{  
i_j\} , \;\{ j_k\}\cdots $ of natural numbers. $m_{i_1 },  
\cdots , m_{i_{d_i } }$ take finite values, but they increase  
in correspondence with $n$ and grows $\to\infty$ as $n\to  
\infty$, and depend on $i$. $d_i$ is the maximal `depth' of the  
composition of $\log$s, or the number of $\log$s, depending on $i$  
and of finite value.  
{\footnote{The power is smaller when $m_1 +m_2 +\cdots +m_{i_1 } 
$ is greater for the same $n$, and when it is also the same and  
$m_1$ is smaller, and when it is also the same and $m_2$ is  
smaller, ..., and so on.} } 
 
As the sum of the shape of $(m)_i$ can always be represented  
as the $\exp$ of the infinite sum of the same shape,  
\beqn (m)_i & = & (\pm )e^{(m)_0 }, \;\; (m)_0:=\log \left (\mbox{sum  
of the finite number of ${e^{(m-1)_i } }$s}\right ) \nonumber \\ 
& = & (m-1)_1 +\log \left ({1\pm e^{(m-1)'_2 } +\cdots } 
\right ), \label{ko26 } \eeqn 
type (g) expansion can in fact be written in only `one term'  
$\exp (m)_{i +1 }$. 
 
Now, for the part of $i\leq 0$ in $(m)_i$, satisfying $0\leq  
k_{(m)_i }$, $\exp (m)_i$ can be written within the shape of $(m) 
_i$ as the composition of $e^z$ and $(m)_i$, Then we can write for  
(I) 
\beqn y=bz +\sum_{n=2 }^\infty a_nz^n\sim +\cdots +\sum_{i<0 } 
(\pm )e^{-b_iz^i\sim\cdots }\cdots +\sum_{j<0 }(\pm )e^{-e^{c_jz^j\sim\cdots }. 
.. }\cdots  \nonumber \\ 
+\sum_{k<0 }(\pm )e^{-e^{e^{d_kz^k\sim\cdots }\cdots } 
\cdots }\cdots , \label{ko27 } \eeqn 
where $b_i, c_j, d_k, \cdots >0$, $\sim$ represents the  
abbreviation of $\log z\sim$, and $\cdots$ the higher order  
terms. The power of $y''/y$ can be classified by whether $b=0$  
or not, and what is the first of $b_i, c_j, d_k, \cdots $ such  
that the corresponding term is not 0: 
\beqn \frac{y'' }{y }\to\left \{ \begin{array}{l}(\pm )\;z^{n-m\epsilon - 
3 }\;\; (b\neq 0\;  
\mbox{and}\;  ^\exists a_n\neq 0, \; n-m\epsilon \geq 2) \\ 
(\pm )\;0\;\; (b\neq 0\; \mbox{and}\; ^\forall a_n=0\; \mbox{and}\;  
 ^\exists b_i\;\mbox{or} \; c_j\; \mbox{or}\; d_k\cdots >0) \\ 
+z^{-2 }\;\; (b=0\; \mbox{and} \;  ^\exists a_n\neq 0) \\ 
+z^{2i\pm 2\epsilon -2 }\;\; (b= ^\forall a_n=0\; \mbox{and} \;  ^\exists  
b_i>0) \\ 
+\infty\;\; (b= ^\forall a_n= ^\forall b_i=0\;  
\mbox{and} \;  ^\exists c_j\; \mbox{or} \; d_k\;  
\mbox{or} \cdots >0)\end{array} \right. ,\label{ko29 } \eeqn 
where $^\forall b_i=0$ means that there is no term in $\sum_{i<0 }$. 
 
After all, $\nu_y\leq -2 +\epsilon , \;\; -1\leq \nu_y$ for (I),  
where $\epsilon$ represents the power like $\log z\sim$. 
 
\vspace{.1 in} 
\noindent (h) It is unclear to me whether there are other cases. 
 
\vspace{.2 in} 
\noindent  (CASE 3) $y(z)$ has a non-isolated essential  
singularity at $z=0$. 
 
\vspace{.1 in} 
\noindent (i) When we allow complex coefficients in (g).  
The discussion above is almost valid in this case, except that  
when $a$ is complex $e^{az }$ shows oscillatory behavior, and  
so $y$ is not monotonic as $z\to 0$ and generally has an accumulation point  
of poles or essential singularities, keeping us away from  
defining $k_y$, $\mu_y$, or $\nu_y$. For example,  
\beqn y=z^5\sin (z^{-1 }) \eeqn  
satisfies the condition of (I) and the term with the  
smallest power in $y$ cancels that of $y''$, yet higher order  
oscillation remains.  
 
\vspace{.1 in} 
\noindent (j) It is unclear to me whether there are other cases.  
In such a case $\nu_y$ would not be clearly physical, even if defined.  
\section{Physical Explanation of the Result } 
The above result is not mathematically perfect, but shows that very  
wide types of functions, only by satisfying the second order  
differential equation, can restrict the behavior of the  
potential. Or physically, if there exists a wave function that  
can be applied to every point of the world, the point of nonzero  
charge should also be included in the domain,  
which determines the shape of a force. 
 
As for the non-commutative character of quantum mechanics,  
type (g) expansion is valid under the special rule that we  
must not decompose an exponential until the end of the calculation.  
Each expansion has several infinite series of different order.  
Having nonzero `radius of convergence', it can be calculated as  
a usual function. Instead, near $z=0$, if we do not obey the  
rule and try to calculate by extracting all the terms below a certain  
order, the result, even if finite, may depend on the arrangement  
of terms. (It is known in mathematics that infinite series that do not 
 converge absolutely do not always converge to a unique value.) 
This implies an interesting non-commutative property.  
 
I notice also that the difficulties caused by point-like  
particles may be absent here. If we assume that the existence of  
an eigen function is more fundamental than that of a potential,  
there can be the region where the potential is not defined (where  
the eigen function is 0). Perhaps this possibility is too subtle  
to be distinguished by experiment, though. The analyticity of  
matter field is not a quantity distinguished by finite times  
of measurement. Conversely, this inevitable ambiguity may be  
the origin of gauge uncertainty\cite{Simon}. 
  
\section{Applications} 
I will comment on possible physical applications of the result. 
The first application is to general relativity, where it shows directly  
that in quantum mechanics, an eigen function and a potential  
obey different transformation rules for a nonlinear coordinate  
transformation.  
In usual quantum mechanics, the special property of an eigen  
function, that its 0-points are of order one and near them it  
behaves like $\sin(x)$, saves the potential from divergence. 
 
Another application is to the general theory of renormalization.  
The above consideration explains why some theories nonrenormalizable  
in the usual sense are partially computable. 
The first example of such case is quantum gravity, where the one  
loop quantum corrections to the Newton potential are determined  
by assuming the Einstein-Hilbert action and the perturbation  
around the flat metric and calculating the effective action\cite{Dono } 
\cite{Doba }\cite{Naka }. 
The result is  
\beqn V(r)=-\frac{Gm_1m_2}{r}\left ({1-\frac{G(m_1+m_2)}{rc^2} 
-\frac{127Gh}{30\pi^2r^2c^3}}\right ),\label{Dono} \eeqn 
where $G, h, c, m_1, m_2$ are respectively the Newton constant,  
Planck constant, the speed of light in a vacuum, and the masses of  
the particles. This naturally contains all the corrections at the distance, 
including the classical relativistic correction. The first term in (\ref{Dono})  
is attractive force and others are repulsive. They correspond to the type (e)  
singularity of the eigen function. Of course, whether or not the assumption  
is valid for very high energy is another story\cite{Wein }, though. 
  
The second example is perturbative QCD, where it is shown that in a confined theory  
the poles and branch points of the true Green functions are generated by the  
Physical hadron states in the unitarity relation, and no singularities related to  
the underlying quark and gluon degrees of freedom should appear \cite{Oehm }. 
Detailed  
discussions about these topics - Borel summation, renormalons, Landau singularities  
- are in \cite{Capr }, so I only mention here that the Callan-Symanzik equation 
\beqn \mu\frac{d}{d\mu}g_\mu =\beta (g_\mu )\eeqn 
just means that the cutoff scale $\mu (g)$ as the function of a coupling constant,  
when differentiated once,  
does not always behave like one with its power smaller by $1$. 
(It seems peculiar that the cutoff scale depends on a coupling constant,  
but I think the idea of multi-valued coupling constant interesting.) 
 
Other applications may include the spherical-symmetric part of the  
effective field equation of the Higgs potential, 
where we can extend the potential to the more general functional of  
scalar field $\phi$ without breaking gauge symmetry.  
\section{Example} 
We can extend the results to dimension $N>1$ as follows. If we 
assume that the eigen function $y$ is a N-dimensional spherical 
symmetric function $R(r)$ (i.e. orbital angular momentum is 0),  
short distance limit behavior (\ref{ko29 }) is clearly replaced by  
\beqn \frac{\Delta R(r) }{R(r)} & = & 
 \frac{R''}{R}+\frac{N-1}{r}\frac{R'}{R} \nonumber \\ 
 & \to & \left \{ \begin{array}{l} +(N-1)r^{-2}\;\; (b\neq 0) \\ 
+n(n+N-2)r^{-2 }\;\; (b=0\; \mbox{and} \;  ^\exists a_n\neq 0) \\ 
+(-ib_i)^2r^{2i\pm 2\epsilon -2 }\;\; (b= ^\forall a_n=0\; \mbox{and} \;  ^\exists  
b_i>0) \\ 
+\infty\;\; (b= ^\forall a_n= ^\forall b_i=0\; \mbox{and} \;  ^\exists  
c_j\; \mbox{or} \; d_k\;  
\mbox{or} \cdots >0)\end{array} \right. .\label{ko30 } \eeqn 
 
We can extend the results to $r\to \infty$ case as follows. If we 
 change the variable to $z:=\frac{1}{r}$ and assume that $R(z)$ is  
$C^2$ class (expanded as below)  
\beqn R=a+bz  & + & \sum_{n=2 }^\infty  
a_nz^n\sim +\cdots +\sum_{i<0 }(\pm )e^{-b_iz^i\sim\cdots } 
\cdots  \nonumber \\ 
 & + & \sum_{j<0 }(\pm )e^{-e^{c_jz^j\sim\cdots }. 
.. }\cdots +\sum_{k<0 }(\pm )e^{-e^{e^{d_kz^k\sim\cdots }\cdots } 
\cdots }\cdots , \label{ko27' } \eeqn 
(\ref{ko30 }) is clearly replaced by 
\beqn \frac{\Delta R(r) }{R(r)} & = & 
 \frac{1}{R(z)}\left \{{\frac{dz}{dr}\frac{d}{dz} 
\left ({\frac{dz}{dr}\frac{dR(z)}{dz}}\right ) 
+(N-1)z\frac{dz}{dr}\frac{dR(z)}{dz}}\right \} \nonumber \\ 
 & = & z^4\frac{R''(z)}{R(z)}-z^3(N-3)\frac{R'(z)}{R(z)} \nonumber \\ 
 & \to & \left \{ \begin{array}{l} (3-N)\frac{b}{a}z^{3}\;\; (a\neq 0\; 
  \mbox{and} \; b\neq 0\; \mbox{and} \;  N\neq 3) \\ 
(n-N+2)n\frac{a_n}{a}z^{n+2}\;\; (a\neq 0\;  \mbox{and} \; b=0\; \mbox{and} \;  
^\exists a_n\neq 0\; \mbox{and} \;  N\neq 3) \\ 
(n-1)n\frac{a_n}{a}z^{n+2}\;\; (a\neq 0\; \mbox{and} \; ^\exists a_n\neq 0\;  
\mbox{and} \;  N=3) \\ 
(\pm )\; 0\;\; (a\neq 0\; \mbox{and} \; b=^\forall a_n=0 \; \mbox{and} \; 
 ^\exists b_i\; \mbox{or} \; c_j\; \mbox{or} \; d_k\; \mbox{or} \cdots >0) \\ 
(3-N)z^{2}\;\; (a=0\; \mbox{and} \; b\neq 0\; \mbox{and} \;  N\neq 3) \\ 
(n-1)n\frac{a_n}{b}z^{n+1}\;\; (a=0\;  \mbox{and} \; b\neq 0\; \mbox{and} \;  
^\exists a_n\neq 0\; \mbox{and} \;  N=3) \\ 
(\pm )\; 0\;\; (a=0\;  \mbox{and} \; b\neq 0\; \mbox{and} \; ^\forall a_n=0\;  
\mbox{and} \; ^\exists b_i\; \mbox{or} \; c_j\; \mbox{or} \; d_k\; \mbox{or} \cdots  
>0\; \mbox{and} \;  N=3) \\ 
(n-N+2)nz^{2}\;\; (a=b=0\; \mbox{and} \; ^\exists a_n\neq 0) \\ 
+(-ib_i)^2z^{2i\pm 2\epsilon +2 }\;\; (a=b= ^\forall a_n=0\; \mbox{and} \;  ^\exists  
b_i>0) \\ 
+\infty\;\; (a=b= ^\forall a_n= ^\forall b_i=0\; \mbox{and} \;  ^\exists  
c_j\; \mbox{or} \; d_k\;  
\mbox{or} \cdots >0)\end{array} \right. .\label{ko31 } \eeqn 
Noting that $2\leq n$ and $i<0$, we conclude that the power of potential 
 $V\to r^\nu$ as $r\to\infty$ is $\nu\leq -3\; ; \;\;-2-\epsilon\leq \nu$. 
There is no reason to assume that $R(z)$ is $C^2$ class, but more 
 natural normalizability condition that $R(r)$ is a $L^2$ function leads  
 to small modifications $a=b=0$ and $N<2n$ (instead of $2\leq n$) in 
 (\ref{ko27' }) and (\ref{ko31 }). Notice also that (\ref{ko30 }) 
for the more general cases of $N, a$ can be obtained from (\ref{ko31 }) 
by the trivial replacement $N\to 4-N$ and $z\to r$ with its power 
smaller by $4$.
 
{\bf Comment on normalizability} \\ 
If $R(r)$ is not a $L^2$ function, that does not always mean  
contradiction. I think that $\delta$ function like sharpness of $R(r)$ 
is not realistic but for $r\to\infty$ the `generalized expectation value'  
of a physical operator $A$ can be defined for $R(r)$ as follows: 
\beqn <A>:=\lim_{L\to\infty}\frac{\int_0^L r^{N-1}dr R^*(r)AR(r)} 
{\int_0^L r^{N-1}dr R^*(r)R(r)}. \label{ko32 } \eeqn 
The above definition may have little physical meaning in case it only  
depends on the value of the field on the surface of a sphere, but  
sometimes, for example when $N=1$ and $R(r)=\sin r$, can be valid.
 
{\bf Comment on uniqueness} \\ 
The solution of the following `2 dimensional weak exterior Dirichlet problem' is not  
unique:\\ 
the function $u(x, y)$ is defined on and on the exterior of the circle  
$x^2+y^2=a^2$ (called C), satisfying the Laplace equation $\Delta u(x, y)=0$ and  
being $0$ on C. Determine $u(x, y)$.\\ 
The proof is as follows. Let $u(x, y)$ be $Re\:\: z-\frac{a^2}{z}$, where $z=x+iy$. 
Because of the Cauchy-Riemann equation, the real part of an analytic function becomes  
automatically harmonic, which shows that $u(x, y)$ is a nontrivial  
solution. It clearly is $0$ on C. In fact we can find infinite number of solutions to  
the problem by the replacement $z\to  
\frac{z^{n+1}}{a^n}$. Furthermore, if we include multi-valued function, another type  
of solutions can be found as follows. 
\begin{enumerate} 
\item Let's take two multi-valued function $f(z)$ and $g(z)$ analytic on and on the  
exterior of C, having a single-valued branch defined outside the cut $z=0, 0\leq Re\; z$. 
\item Combine them so as not to have a gap at the cut on C. For example, define a  
new function  
$h(z):= \Delta_g f(z)-\Delta_f g(z)$, where $\Delta_f:= f(ae^{+i0})- f (ae^{i(2\pi  
-0)}), \Delta_g:= g(ae^{+i0})-g(ae^{i(2\pi -0)})$ are the gap of $f(z), g(z)$. 
\item As the branch $Re\; h(z), z=re^{i\theta}, 0\leq\theta\leq 2\pi$ is continuous  
on C and is $0$ at $z=a$, there is the unique sine Fourier expansion of it on C,  
i.e., $Re \:\: h(ae^{i\theta})=\sum_{n=0}^\infty h_n\sin (\frac{n\theta}{2})$. 
\item There is the corresponding function $\hat h(z):=\sum_{n=0}^\infty  
-h_n(\frac{a}{z})^{\frac{n}{2}}$ such that $Im\; \hat h(z)=\sum_{n=0}^\infty  
h_n\sin (\frac{n\theta}{2})$ on C. 
\item Define $\hat H(x, y):= Re\; h(z)-Im\; \hat h(z)$, where $z:=x+iy$. Then, 
the branch  
$\hat H(x, y)$ can be a nontrivial solution to the problem for, for example,  
$f(z)=\frac{1}{\log z}, g(z)=\frac{1}{z\log z }$. Notice that $\hat H(x, y)$ is  
generally multi-valued on the cut except for the point $(a, 0)$. 
\end{enumerate} 
 
{\bf Comment on relativistic effect and the sign of potential} \\ 
For the time-independent and spherical-symmetric $U(1)$ gauge field $A^\mu =(\phi  
(r), 0, 0, 0)$, (\ref{ko31 }) becomes an exact relativistic Schr\"{o}dinger equation  
by the replacement  
$V(r)\to\frac{m^2c^4-(E-e\phi )^2}{{\bar h}^2c^2}$, where $E, e, m$ are respectively  
dimensionthe energy, the charge, and the mass of the spherical-symmetric scalar field  
$R(r)$\cite{Schiff}. Of course, fermion field equation is another story. 
Above results show that for a physical 
dimension $N=1, 2, 3$, the sign of a potential $V$ must be positive for
$\nu\leq -2+\epsilon\;\; (r\to 0)$ and $-2-\epsilon\leq \nu\;\; (r\to\infty)$
, but can be negative for other cases.
\section{Conclusion } 
In this paper we classified possible singularities of a potential for the  
one dimensional Schr\"{o}dinger equation, assuming that a  
potential $V$ has at least one $C^2$ class eigen function. The result  
crucially depends on the analytic property of the eigen function near its  
0 point. We also discussed the extension to dimension $N$ and the long
distance limit. There are interesting applications to quantum gravity
and to the general theory of renormalization. 
\section*{Acknowledgements} 
I am grateful to Izumi Tsutsui and Toyohiro Tsurumaru for  
useful discussion. I also appreciate Tsutomu Yanagida and  
Ken-ichi Izawa for reading the manuscript.


\begin{thebibliography}{99}
\bibitem{uniq} All the terms are arranged in the order of ascending powers. For example, $y(r)= 7r^{\frac{73}{95}}+5r^{\sqrt 5}+
3r^{\pi}+r^{2\sqrt 5}+3r^{\frac{77}{5}}$. Then, there is the natural unique analytic continuation 
of $y(r)$ from the real axis to the complex region around $z=0$ without the cut $z\leq 0$.
\end{thebibliography}
\end{document}